\documentclass[aps,pra,twocolumn,amsmath,amssymb,superscriptaddress,longbibliography]{revtex4-1}
\usepackage[english]{babel}
\usepackage{amssymb}
\usepackage{dcolumn}
\usepackage{bm}
\usepackage{graphicx}
\usepackage{amsmath}
\usepackage{graphicx}        
\graphicspath{{pict/}{}}

\usepackage{dcolumn}
\usepackage{bm}
\usepackage[pdfstartview=Fit, CJKbookmarks=true, bookmarksnumbered=true, bookmarksopen=true, colorlinks=true, pdfborder=001, citecolor=blue, linkcolor=blue, urlcolor=blue, linktocpage=true] {hyperref}

\begin{document}

\title{Uncover band topology via quantized drift in two-dimensional Bloch oscillations}

\author{Bo Zhu}
\affiliation{Institute of Mathematics and Physics, Central South University of Forestry and Technology, Changsha 410004, China}

\author{Shi Hu}
\affiliation{School of Optoelectronic Engineering, Guangdong Polytechnic Normal University, Guangzhou, 510665, China}

\author{Honghua Zhong}
\altaffiliation{hhzhong115@163.com.}
\affiliation{Institute of Mathematics and Physics, Central South University of Forestry and Technology, Changsha 410004, China}

\author{Yongguan Ke}
\altaffiliation{keyg@mail2.sysu.edu.cn.}
\affiliation{Guangdong Provincial Key Laboratory of Quantum Metrology and Sensing $\&$ School of Physics and Astronomy, Sun Yat-Sen University (Zhuhai Campus), Zhuhai 519082, China}

\date{\today}

\begin{abstract}
We propose to measure band topology via quantized drift of Bloch oscillations
in a two-dimensional Harper-Hofstadter lattice subjected to tilted fields in both directions.
When the difference between the two tilted fields is large, Bloch oscillations uniformly sample all momenta, and hence the displacement in each direction tends to be quantized at multiples of the overall period, regardless of any momentum of initial state.
The quantized displacement is related to a reduced Chern number defined as a line integral of Berry curvature in each direction, providing an almost perfect measurement of Chern number.
Our scheme can apply to detect Chern number and topological phase transitions not only for the energy-separable band, but also for energy-inseparable bands which cannot be achieved by conventional Thouless pumping or integer quantum Hall effect.
\end{abstract}

\maketitle

\section{Introduction\label{SSec1}}

Topological band theory~\cite{bansil2016colloquium} provides a general framework for exploring a variety of topological states and phenomena such as integer quantum Hall effects~\cite{klitzing1980new,thouless1982quantized}, topological insulators~\cite{Kane2005Quantum, Bernevig2006Quantum, konig2007quantum, Chang2013Experimental,Xu2014Observation}, and Thouless pumping~\cite{thouless1983quantization,ke2016topological,hu2019dispersion,Lin2020}.
A key notion is topological invariant (e.g., Chern number and winding number) which characterizes the robust global property of Bloch states of certain energy bands.
Topological phase transition happens when a topological invariant changes.
Extending the topological states from solid-state materials to ultracold atoms in optical lattices have attracted tremendous interests in recent years~\cite{Aidelsburger2013Realization, miyake2013realizing, jotzu2014experimental, 2015Wavepacket}.
%
%
Ultracold atomic system, as an excellent platform for quantum simulations, provides unprecedented opportunities for studying topological states, in particular, uncovering band topology.

Many schemes were proposed to explore band topology and some of them have been realized in experiments of ultracold atoms.
In momentum space, the bulk topological invariants can be determined by linking numbers~\cite{wang2017scheme,tarnowski2019measuring}, band-inversion surfaces~\cite{sun2018uncover,zhang2018dynamical,zhang2019dynamical} in quench dynamics, and dynamic winding number~\cite{zhuDynamic2020} based on time-averaged spin textures.
However, these dynamical methods need to measure different observables involving all momentum states at different time, which turn out to be redundancy and inefficient.
In real space, winding number can be measured by a mean chiral displacement in quantum walks~\cite{zeuner2015observation,Longhi2018Probing,2019Probing}, and Chern number can be extracted by a mean displacement in Thouless pumping or transverse velocity in the integer quantum Hall effect~\cite{ke2016topological,hu2019dispersion}.
However, these methods in real space require that the initial state uniformly occupies all the momentum states of a band, which in general is a challenge.
It becomes a dilemma about either suffering labored measurements or overcoming difficulty in preparing an input state.

The above dilemma may be solved by utilizing Bloch oscillations under external tilted fields, which have been proved to be a powerful method for exacting various geometric features of Bloch bands, including Berry phase~\cite{atala2013direct, Duca2015An}, Berry curvature~\cite{Price2012,jotzu2014experimental, Martin2017Experimental} and Chern number~\cite{aidelsburger2015measuring, 2019Imaging, 2020Probing, Kepumping2020}.
In particular,  Berry curvature was proposed to be mapped from semiclassical dynamics by  using two-dimensional Bloch oscillations (i.e., applying external forces in two dimensions)~\cite{Price2012}.
However, to further extract Chern number, this protocol needs to measure the displacement difference under positive and negative forces to cancel the contribution of group velocity from energy dispersion, and then to average the displacement difference under different initial momenta.
Recently, we have shown how to directly extract Chern number via a one-dimensional quantized topological pumping assisted by Bloch oscillations that does not require uniform band occupation~\cite{Kepumping2020}. This is because Bloch oscillations uniformly sample all the momenta.
It is highly nontrivial to directly and efficiently measure band topology via Bloch oscillations in two dimensions.

In this paper, we reveal a quantized drift of Bloch oscillations as a \emph{direct measurement} of Chern number in a two-dimensional Harper-Hofstadter-like lattice subjected to external fields in both $x$ and $y$ directions. Such scheme is readily accessible  with typical experimental settings of ultracold atomic gases in optical lattices~\cite{Aidelsburger2013Realization,aidelsburger2015measuring}.
We give analytical expressions for the mean displacement in both $x$ and $y$ directions via the adiabatic transport theorem.
As the ratio between external fields in $x$ and $y$ directions approaches to $0$ or $\infty$, we find that the mean displacement in each direction is almost perfectly quantized and independent of the initial momentum.
%
The quantized displacement in each direction is given by a time integral of Berry curvature dubbed as reduced Chern number (RCN), which is directly related to the conventional Chern number.
According to the quantized drift in the overall period of Bloch oscillations, one can successfully detect topological phase transitions and Chern number for an energy-separable band.
With the increase of ratio of tilts in two directions, the accuracy of Chern number near the phase transition point will be gradually improved.
Importantly, we show a direct measurement of the Chern number for an energy-inseparable band via RCN, which can never be achieved by the conventional Hall-response scheme with tilt in only one direction~\cite{aidelsburger2015measuring}.
Compared with Ref.~\cite{Price2012}, our method has several advantages listed as follows. (i) Our protocol automatically cancels the group velocity from energy dispersion, and directly gives the Chern number via displacement in a single wavepacket dynamics, in which there is no need to inverse the force to eliminate the contribution of displacement from energy dispersion;
(ii) We also reveal that the quantized drift in Bloch oscillations is independent of the momentum of the initial Gaussian wavepacket, reducing difficulty of the initial state preparation and many wavepacket dynamics with different initial momenta;
(iii) We give a direct measurement of super-band topology which has never been discussed in Ref.~\cite{Price2012}.

The rest of the paper is organized as follows.
In Sec.~\ref{SSec2}, we give a physical description of the model.
In Sec.~\ref{SSec3},
we give the definition of RCN and establish the relation between RCN and the quantized  displacement  in two-dimensional Bloch oscillations.
In Sec.~\ref{SSec4}, we show how to use Bloch oscillations to extract Chern numbers for an energy-separable band
in the subsection~\ref{energy-separable} and for an energy-inseparable band in subsection~\ref{super-band}.
At last, we give a conclusion and discussion in Sec.~\ref{SSec5}.

\section{Tilted Harper-Hofstadter lattice \label{SSec2}}
\begin{figure}[!htp]
	\center
	\includegraphics[width=0.9\columnwidth]{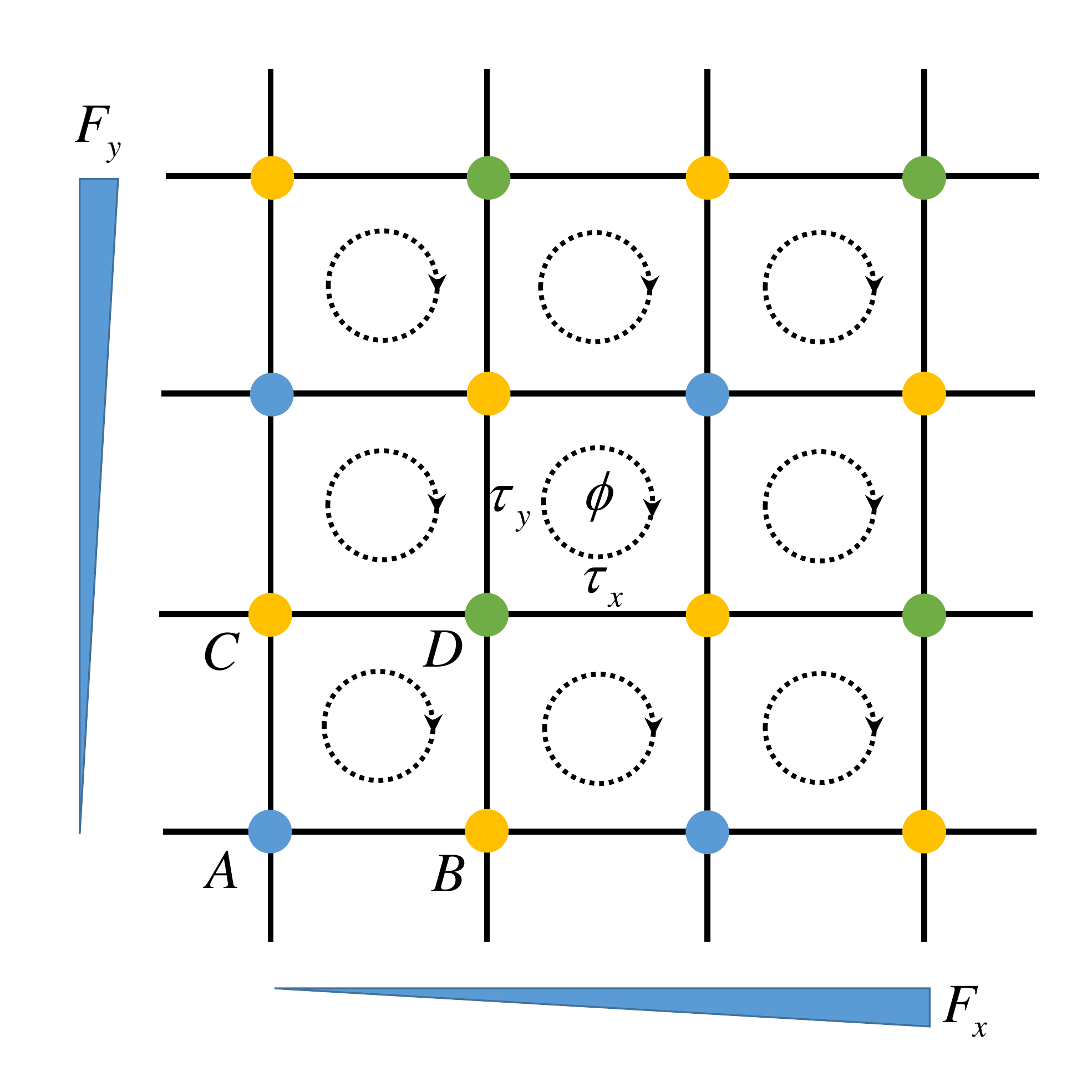}
	\caption{Schematic diagram of the Hofstadter-like optical lattice with additional tilts along $x$ and $y$ directions.
		$\tau_{x(y)}$ and $F_{x(y)}$ are the hopping strength and tilted strength along $x(y)$ direction,
		$(A, B, C, D)$ denote the sublattices in a unit cell with onsite energies $(-\delta,0,0,+\delta)$, and $\phi$ is the accumulated phase per plaquette due to the gauge field.
	}\label{Fig1}
\end{figure}

We consider a single particle in a two-dimensional tilted superlattice in the presence of a uniform flux per plaquette, see Fig.~\ref{Fig1}.
The corresponding Hamiltonian consists of three parts,
\begin{eqnarray} \label{Ham_1}
H=H_1+H_2+H_3,
\end{eqnarray}
with
\begin{eqnarray}
H_1&=&-\sum_{m, n}\tau_{x} c_{m+1, n}^{\dagger} c_{m, n}+\tau_{y} e^{i 2 \pi \beta m} c_{m, n+1}^{\dagger} c_{m, n}+\mathrm{h.c.}, \nonumber \\
H_2&=&-\sum_{m, n}\frac{\delta}{2}[(-1)^m+(-1)^n] c_{m, n}^{\dagger} c_{m, n}, \nonumber \\
H_3&=&-\sum_{m, n}(F_x m+F_y n) c_{m, n}^{\dagger} c_{m, n}.
\end{eqnarray}
Here, $H_1$ is the conventional Harper-Hofstadter Hamiltonian~\cite{Douglas1976Energy,2005Fractional,2020Detecting,2020Fractional},
and the Peierls phase accounts for the presence of a flux $\phi=2 \pi \beta$ per plaquette.
$c_{m, n}^{\dagger}(c_{m, n})$ creates (annihilates) a particle at site $(m, n)$, $\tau_{x}$ and $\tau_{y}$ are the hopping strengths along $x$ and $y$ directions.
$H_2$ describes a staggered detuning $\delta$ between sublattices in two directions,
which can be used to induce topological phase transition~\cite{aidelsburger2015measuring}.
When $\tau_{x}=\tau_{y}=\tau$, $|\delta|\ge 2 \tau$ and $|\delta|<2 \tau$ correspond to topologically trivial and nontrival phases, respectively.
$H_3$ represents the tilts of the square lattice,
where $F_x$ and $F_y$ are the tilted strengths along $x$ and $y$ directions,
which may be realized by applying a magnetic field gradient or subjecting the superlattice along the gravity with an angle.
Because the tilted strengths are very weak compared to the coupling strengths,
$H_3$ can be treated as a perturbation~\cite{1996Berry},
which means that $H_3$ does not destroy the topological properties of the system.

The time evolution of state $|\psi(t)\rangle$ can be obtained by solving the Schr\"{o}dinger equation
$i \hbar \frac{\partial}{\partial t} |\psi(t)\rangle=H|\psi(t)\rangle$.
Hereafter we set $\hbar=1$ for simplicity. By making a unitary transformation $|\tilde{\psi}(t)\rangle=\exp(iH_3t)|{\psi}(t)\rangle$, we can equivalently deal with the problem in a rotational framework, and  $|\tilde{\psi}(t)\rangle$ is governed by
\begin{equation}
i \frac{\partial}{\partial t} |\tilde\psi(t)\rangle=H_{rot}(t)|\tilde \psi(t)\rangle
\end{equation}
with the time-dependent Hamiltonian $H_{rot}(t)=e^{iH_3t}(H_1+H_2)e^{-iH_3t}$, which is explicitly given by
\begin{eqnarray} \label{Ham_2}
&&H_{rot}(t)=-\frac{\delta}{2}\sum_{m, n}[(-1)^m+(-1)^n] c_{m, n}^{\dagger} c_{m, n}-  \\
&&\sum_{m, n}\tau_{x}e^{iF_xt} c_{m+1, n}^{\dagger} c_{m, n}+\tau_{y} e^{iF_yt} e^{i 2 \pi \beta m}c_{m, n+1}^{\dagger} c_{m, n}+\mathrm{h.c.} \nonumber
\end{eqnarray}
If $\beta=p/q$, where $p$ and $q$ are co-prime numbers, the Hamiltonian has magnetic translational symmetry.
In the whole paper, we consider $\beta=1/4$ without loss of generality. Under the periodic boundary condition, one can obtain the Hamiltonian in quasimomentum space $H_{r o t}(t)=\sum_{k_x,k_y} \Psi_{{k_x,k_y}}^{\dagger} h(k_x,k_y, t) \Psi_{{k_x,k_y}}$ with
\begin{eqnarray} \label{Ham_3}
h=2\left(\begin{array}{cccc}
-\frac{\delta}{2} & -\tau_{y}\cos \kappa_{y} & -\tau_{x}\cos \kappa_{x} & 0 \\
-\tau_{y}\cos \kappa_{y} & 0 & 0 & \tau_{x}\sin \kappa_{x} \\
-\tau_{x}\cos \kappa_{x} & 0 & 0 & i \tau_{y}\sin \kappa_{y} \\
0 & \tau_{x}\sin \kappa_{x} & -i \tau_{y}\sin \kappa_{y} & \frac{\delta}{2}
\end{array}\right). \nonumber \\
\end{eqnarray}
Here, $\Psi_{\textit{\textbf{k}}}=\left(c_{\textit{\textbf{k}}, A}, c_{\textit{\textbf{k}}, B}, c_{\textit{\textbf{k}}, C}, c_{\textit{\textbf{k}}, D}\right)^{T}$, where  $c_{\textit{\textbf{k}}, \alpha}$ destroys a plane wave with quasimomentum $\textit{\textbf{k}}\equiv(k_x,k_y)$ at $\alpha$ sublattice, $\kappa_{x(y)}=k_{x(y)}-F_{x(y)} t$ and $k_{x(y)}\in[-\pi/2,\pi/2]$ (see Appendix~\ref{Appendix1} for details).
We can numerically obtain the eigenvalues $E_l(k_x, k_y, t)$ and eigenstates $ |\mu_l(k_x,k_y, t)\rangle$ by diagonalizing the Hamiltonian, $h(k_x,k_y, t)|\mu_l(k_x,k_y, t)\rangle = E_l(k_x, k_y, t) |\mu_l(k_x,k_y, t)\rangle$, where $l$ is the band index.

\begin{figure}[!htp]
\center
\includegraphics[width=\columnwidth]{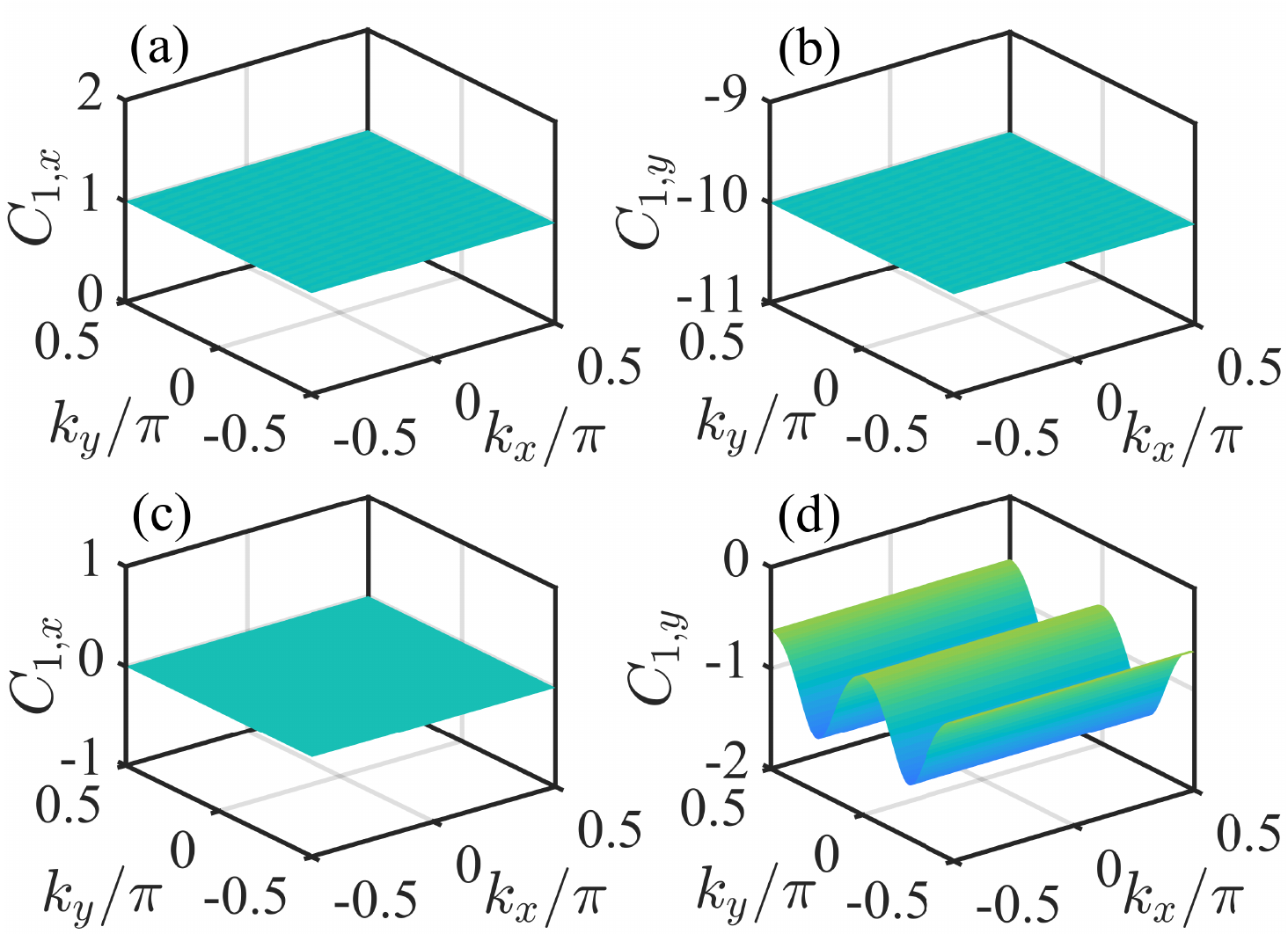}
\caption{Reduced Chern number of the first band as a function of $(k_x,k_y)$. (a) $C_{1,x}(k_{x}, k_{y})$ and (b) $C_{1,y}(k_{x}, k_{y})$ for tilts in both $x$ and $y$ directions ($F_x/F_y=\eta_x/\eta_y=10$, $F_x=0.04, F_y=0.004$).
	(c) $C_{1,x}(k_{x}, k_{y})$ and (d) $C_{1,y}(k_{x}, k_{y})$ for tilts in only $x$ direction ($F_x=0.0402, F_y=0$). The other parameters are chosen as $\tau_{x}=\tau_{y}=0.2$, and $\delta=0$.
}\label{Fig2}
\end{figure}

\section{Reduced Chern number \label{SSec3}}
Consider that the initial state is prepared as a Bloch state with quasimomentum ($k_x, k_y$) in the $l$th band.
If the tilted strengths are so weak then the particle adiabatically follows the trajectory $[\kappa_{x}(t),\kappa_y(t)]$ in the $l$th band.
According to the theorem of adiabatic transport~\cite{Xiao2010Berry, 2012Quantized}, the group velocity of the particle along the $x(y)$ direction comes from the energy dispersion and the Berry curvature,
\begin{equation} \label{Ham_4}
v_{l,x(y)}(k_x, k_y, t)=\frac{\partial E_{l}(k_x, k_y, t)}{ \partial k_{x(y)}}+\mathcal{F}_{l,x(y)}(k_x, k_y, t),
\end{equation}
with the Berry curvature given by
\begin{equation}
\mathcal{F}_{l,x(y)}=-2 \operatorname{Im}\sum_{l^{\prime} \neq l} \frac{\left\langle u_{l}\left|\partial_{k_{x(y)}} h \right| u_{l^{\prime}}\right\rangle\left\langle u_{l^{\prime}}\left|\partial_{t} h \right| u_{l}\right\rangle}{\left(E_{l}-E_{l^{\prime}}\right)^{2}}.
\end{equation}
Because the energy is a (quasi-)periodic function of time, if the band topology is completely trivial, the first term of group velocity will periodically oscillate and hence induce the conventional Bloch oscillations in both directions.
In the following subsection, we will show how the nontrival Berry curvature induces quantized drift in the Bloch oscillations.

\subsection{Rational case}
We first consider the case that the tilts in two directions
are nonzero and commensurate $F_x/F_y=\eta_x/\eta_y$, where $\eta_x$ and $\eta_y$ are coprime integers.
If $F_x=F_y$, the period of Bloch oscillations in $x$ direction $T_x=2\pi/F_x$ is the same as that in $y$ direction $T_y=2\pi/F_y$.
However, if $F_x \ne F_y$, $T_x$ and $T_y$ may be not the period in their own directions.
The coexistence of the tilts in two directions leads to an overall period $T_{o}=\eta_x T_x=\eta_y T_y$.
%

For a Bloch state with quasimomentum $(k_{x}, k_{y})$, the mean displacement in $x$ and $y$ directions $(\Delta X,\Delta Y)$  at time $t$ can be given by the semi-classical expression
\begin{eqnarray} \label{Ham_6}
\Delta X(k_{x}, k_{y}, t)&=&\int_{0}^{t} v_{l,x}(k_{x}, k_{y}, t') d t', \nonumber \\
\Delta Y(k_{x}, k_{y}, t)&=&\int_{0}^{t} v_{l,y}(k_{x}, k_{y}, t') d t'.
\end{eqnarray}
In the commensurate case, because the energy bands are periodic functions of time,
the integral of dispersion velocity is exactly zero in the overall period.
Due to the nontrivial Berry curvature, the anomalous group velocity thus plays a determinant role in the mean displacement.
We define the time integral of Berry curvature over the overall period as
\begin{equation} \label{Ham_8}
C_{l,x(y)}(k_{x}, k_{y})=\frac{1}{ q} \int_{0}^{T_{o}} \mathcal{F}_{l,x(y)}(k_{x},k_{y}, t) d t,
\end{equation}
When $\eta_x/\eta_y$ approaches $0$ or $\infty$,  $C_{l,x}(k_{x}, k_{y})$ and $C_{l,y}(k_{x}, k_{y})$ respectively tend to be integer numbers $C_{l,x}^0$ and $C_{l,y}^0$, independent of the initial momentum value of a Bloch state (see Appendix~\ref{Appendix4} for details).
Interestingly, the integer numbers $C_{l,x}^0$ and $C_{l,y}^0$ are related to the conventional Chern number of the $l$th band  (see Appendix~\ref{Appendix2} for details),
\begin{eqnarray} \label{Ham_7}
C_{l} = C_{l,x}^0/\eta_y=-C_{l,y}^0/\eta_x.
\end{eqnarray}
For this reason, we name the line integral of Berry curvature $C_{l,x(y)}$ as reduced Chern number.
Then the displacement in an overall period tends to be quantized values
\begin{eqnarray} \label{Ham_10}
\Delta X(k_{x}, k_{y}, T_{o})&=& q C_{l,x}(k_{x}, k_{y}), \nonumber \\
\Delta Y(k_{x}, k_{y}, T_{o})&=& q C_{l,y}(k_{x}, k_{y}).
\end{eqnarray}
Combining Eqs.~\eqref{Ham_7} and ~\eqref{Ham_10}, we find that the mean displacement in $x$ and $y$ directions will satisfy
\begin{equation}
\mathop {\lim }\limits_{{\eta_x(\eta _y)} \to \infty } (\Delta X(T_{o})\eta_x+\Delta Y(T_{o})\eta_y) \rightarrow 0.
\end{equation}

To verify the above analytical results, we respectively show
$C_{1,x}$ and $C_{1,y}$ as functions of $(k_x, k_y)$ in Figs.~\ref{Fig2}(a) and \ref{Fig2}(b), with the parameters  $F_x/F_y=\eta_x/\eta_y=10$, $F_x=0.04, F_y=0.004$ , $\tau_{x}=\tau_{y}=0.2$ and $\delta=0$.
When the tilts in both $x$ and $y$ directions are present and their ratio is small or large enough,  both $C_{1,x}$ and $C_{1,y}$ are extremely flat for all momenta and close to quantized values, $C_{1,x}^0=1$ and $C_{1,y}^0=-10$, respectively.
Considering that the Chern number $C_1=1$, it is easy to find that the formula~\eqref{Ham_7} is satisfied.
The fact that $C_{1,x}/\eta_y$ and $C_{1,y}/\eta_x$ are almost independent of the initial momentum means that one can use a single momentum state to detect Chern number.
This is strikingly different from the case that a tilt is present in only $x$ direction, see Figs.~\ref{Fig2}(c) and \ref{Fig2}(d), where
$F_x=0.0402, F_y = 0$ and the overall period becomes $T_{o}= T_x$.
$F_x$ and $F_y$ have the same norm $\sqrt{F_x^2+F_y^2}$ as that in Fig.~\ref{Fig2}(a) and \ref{Fig2}(b).
$C_{1,x}$ is exactly zero, while $C_{1,y}$ maintains the same along $k_x$ direction for a fixed $k_y$ but periodically varies along the $k_y$ direction.
This is because the Berry curvature $\mathcal F_{1,x}=0$ at any time and $\mathcal F_{1,y}$ is a periodic function of $k_y$.

Obviously, in the case of tilt in a single direction, we cannot use an arbitrary single momentum state to detect Chern number.
However, the average of $C_{1,x}(k_{x}, k_{y})$ over $k_y$ is still quantized,
i.e., $\frac{1}{\pi}\int_{-\pi/2}^{\pi/2} C_{1,y}(k_{x}, k_{y}) d k_y=1$,
which is consistent with the first band Chern number $C_1=1$.
This means that if the initial state is an equal superposition
of momentum states involving all $k_y$, the Chern number of the band can also be detected
by mean displacement for the system with the tilt in a single direction.
It is worth noting that the experimental scheme
of detecting Chern numbers of Hofstadter bands for
the tilt in a single direction has been realized~\cite{aidelsburger2015measuring},
where it requires that the topological bands are very flat.
The essential reason is that the flat band makes it easier to uniformly fill the band.

\begin{figure}[!htp]
	\center
	\includegraphics[width=\columnwidth]{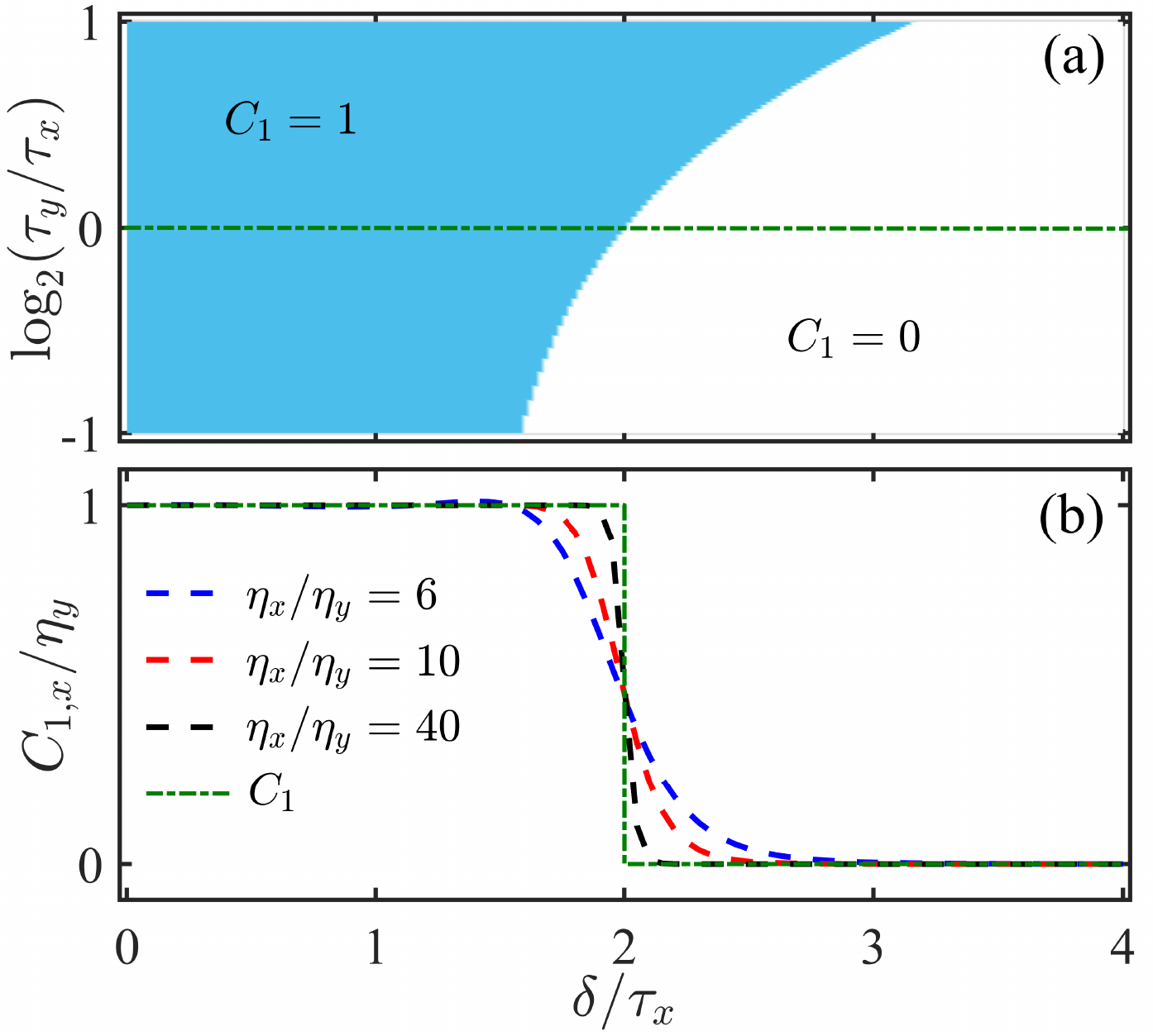}
	\caption{(a) Topological phase diagram.
		(b) RCN as a function of $\delta$ for different commensurate ratios $\eta_x/\eta_y$ (fixed $\sqrt{F_x^2+F_y^2}=0.0402$ and $k_x=k_y=0$), the other parameters are chosen as $\tau_{x}=\tau_{y}=0.2$, corresponding to the green dashed line in (a).
	}\label{Fig3}
\end{figure}

In contrast to previous schemes with tilt in a single direction, according to the definition of RCN,
our proposed scheme with tilts in two directions has greater advantages for experimental detection of Chern number.
Firstly, this scheme does not require a flat topological band;
Secondly, this scheme also does not require the initial states being equal superposition of  all momentum states,
Therefore, RCN may provide an experimentally-friendly way for detecting the topological phase transition.
To this end, we first show the topological
phase diagram in the parameter plane $(\tau_{y}, \delta)$ by using
conventional Chern number, see Fig.~\ref{Fig3}(a).
The Chern numbers of the first band
are $C_1= 1$ and $0$ in the blue and white regions,
where the green dashed-dot line corresponds to the case
of hopping constant $\tau_{x}=\tau_{y}$ and the phase transition point is $\delta/\tau_{x}=2$.
Along the green dashed-dot line,
we give the RCN $C_{1,x}^0/\eta_y$ as a function
of $\delta$ for different ratios of tilts $\eta_x/\eta_y$ in Fig.~\ref{Fig3}(b).
The RCNs are consistent with the first band Chern number $C_1$ away from the phase transition point.
As the ratio of tilts increases, the change of RCN around $\delta/\tau_x=2$ becomes sharper, marking the position of the transition point.
Therefore, the RCN can apply to detect the topological phase transition by choosing $\eta_x/\eta_y\rightarrow \infty$, or equivalently $\eta_x/\eta_y\rightarrow 0$.

\subsection{Irrational case}
\begin{figure}[!htp]
	\center
	\includegraphics[width=\columnwidth]{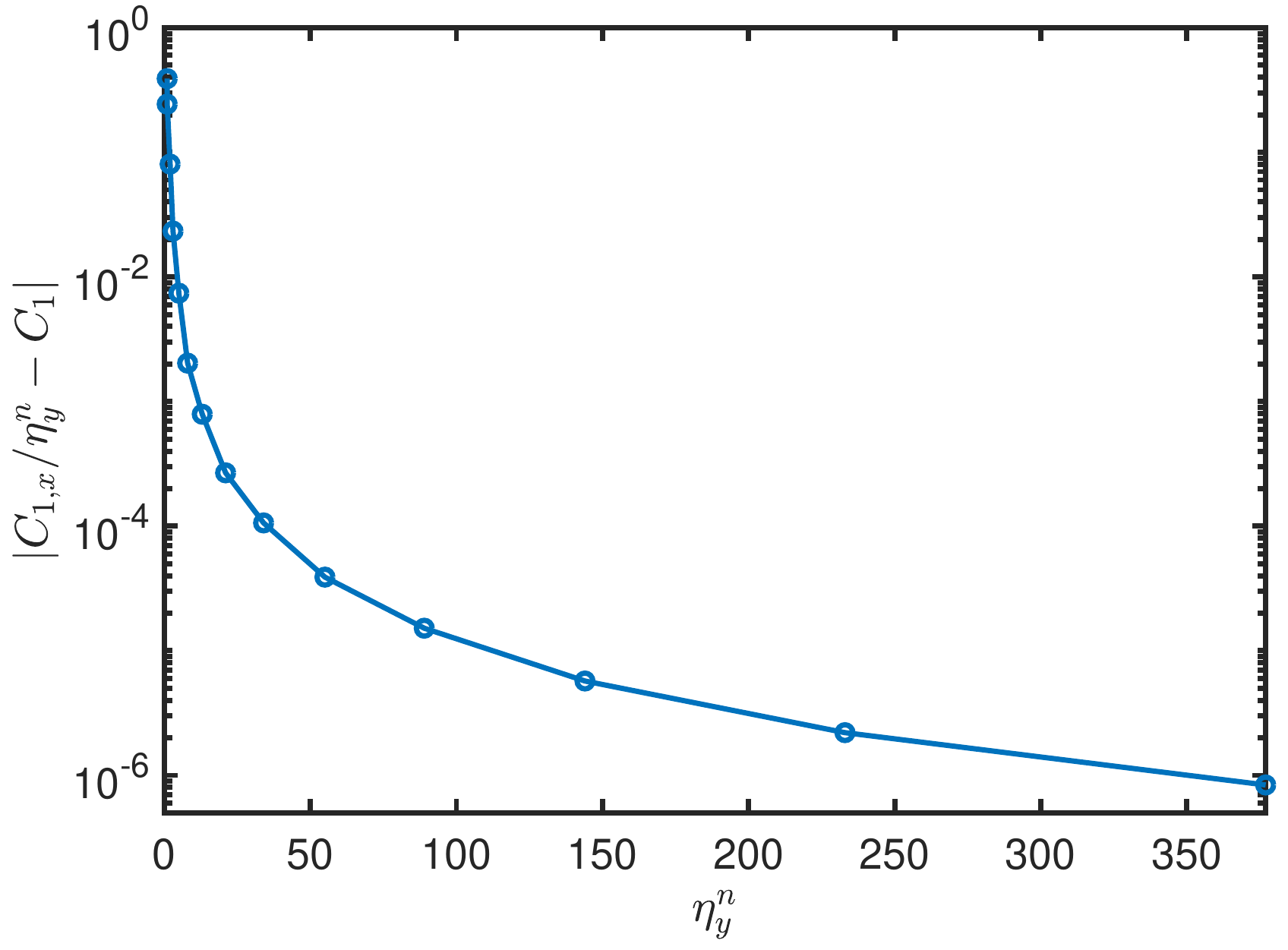}
	\caption{Derivation between $C_{1,x}/\eta_y^n$ and $C_1$ as  $\eta_y^n$ increases for the irrational $F_x/F_y=(\sqrt{5}+1)/2$.
The other parameters are chosen as $F_y=0.004, \tau_x=\tau_y=0.2$, $\delta=0$ and $k_x=k_y=0$.	}\label{Fig4}
\end{figure}

When the tilts in two directions are incommensurate, both the energy spectrum and Berry curvature are quasi-periodic function of time, and hence  there is no such an overall period as the rational case.
However, we can use a sequence of rational ratios $\{\eta_x^n/\eta_y^n\}$ to approach the irrational ratio $F_x/F_y$, and treat the irrational case in a similar way as the rational case.
When $\eta_y^n$ tends to infinite integer, we can reach the RCN and the relation in Eq.~\eqref{Ham_10} still holds.
Taking $F_x/F_y=(\sqrt{5}+1)/2$ as example, we can use a continued fraction representation for the golden ratio, which is given by $\{\eta_x^n/\eta_y^n\}=(1/1,2/1,3/2,5/3,8/5,...,F_{j+1}/F_j,...)$ with the Fibonacci sequence $F_j$.
We show how the derivation between RCN and the conventional Chern number changes with the increase of $\eta_y^n$ in Fig.~\ref{Fig4}.
As expected, the derivation quickly decays as $\eta_y^n$ increases, indicating that the irrational case can be practically treated as the rational case with large $\eta_y^n$.

\section{Two-dimensional quantized-drift Bloch oscillations \label{SSec4}}
In the previous section, we have shown how the Chern number of a given band is related to the quantized drift in the Bloch oscillations based on semiclassical analysis.
To verify such relation, in this section we study the quantum dynamics of a Gaussian wavepacket centered in $(m_0,n_0)$ with arbitrary mean quasimomentum $(k_{x_0},k_{y_0})$ in the $l$th band.
The initial wavefunction at the site $(m, n)$ hence is given by
\begin{eqnarray} \label{Ham_11}
\psi_{m,n}(0)=\zeta e^{-\frac{(m-m_0)^2+(n-n_0)^2}{4 D^2}} \widetilde{\mu}_{m,n}^l(k_{x_0},k_{y_0}) e^{i(k_{x_0} m+k_{y_0} n)}, \nonumber
\end{eqnarray}
where $\zeta$ is a normalization factor, $D$ is the initial wavepacket width, $\widetilde{\mu}_{m,n}^l(k_{x_0},k_{y_0})$ is the amplitude of the real-space representation of the Bloch state $|\mu_{l}(k_{x_0},k_{y_0},0)\rangle$.
Such type of wavepacket can be prepared by applying an additional harmonic trap~\cite{Lu2016Geometrical}.
In the following calculations, the time evolution of the wavepacket is given by $|\psi(t)\rangle=\exp(-i H t)|\psi(0)\rangle$ according to the Hamiltonian Eq.~\eqref{Ham_1}.
One may alternatively calculate the wavepacket dynamics in the rotating framework with time-dependent Hamiltonian, which gives the same density distribution but increases the calculation resources due to the update of Hamiltonian at each time step.
We then examine the density distribution profile $|\psi_{m,n}(t)|^2$ of the time-evolving wavepacket $\psi_{m,n}(t)$, and the two components of mean displacement in $x$ and $y$ directions
\begin{eqnarray} \label{Ham_12}
\Delta X(t)&=&X(t)-X(0), \nonumber \\
\Delta Y(t)&=&Y(t)-Y(0),
\end{eqnarray}
where $X(t)=\sum_{m,n} m\left|\psi_{m,n}(t)\right|^{2}$ and $Y(t)=\sum_{m,n} n\left|\psi_{m,n}(t)\right|^{2}$.
The dynamical results are shown in subsection~\ref{energy-separable} for the energy-separable band and in subsection~\ref{super-band} for the energy-inseparable band.

\subsection{Direct measurement of separable-band topology} \label{energy-separable}
\begin{figure}[!htp]
	\center
	\includegraphics[width=\columnwidth]{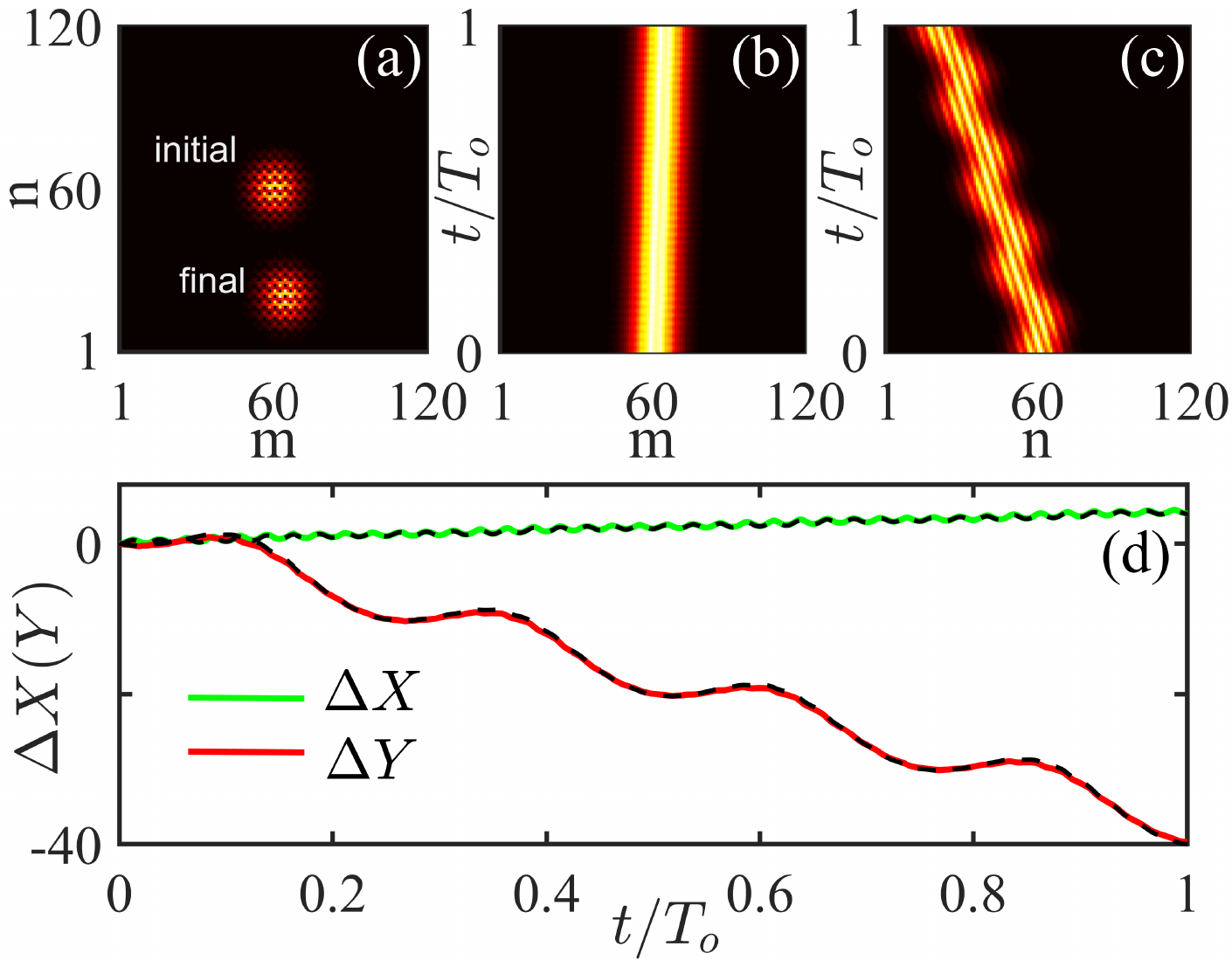}
	\caption{Wavepacket dynamics in topologically nontrivial phase.
		(a) Density distribution of initial states $(t=0)$ and final states $(t=T_{o})$ in real space.
		(b) and (c) correspond to the density evolution project to $x$ and $y$ directions, respectively.
		(d) Drift $\Delta X$ and $\Delta Y$ versus time $t$.
		The green and red solid lines are obtained from quantum dynamics, and the black dashed line is obtained via Eq.~\eqref{Ham_6}.
		The other parameters are chosen as $\tau_{x}=\tau_{y}=0.2$, $D=6$, $\eta_x/\eta_y=10$, $F_x=0.04, F_y=0.004$, $\delta=0$ and $k_{x_0}=k_{y_0}=0$.
	}\label{Fig5}
\end{figure}

\begin{figure}[!htp]
	\center
	\includegraphics[width=\columnwidth]{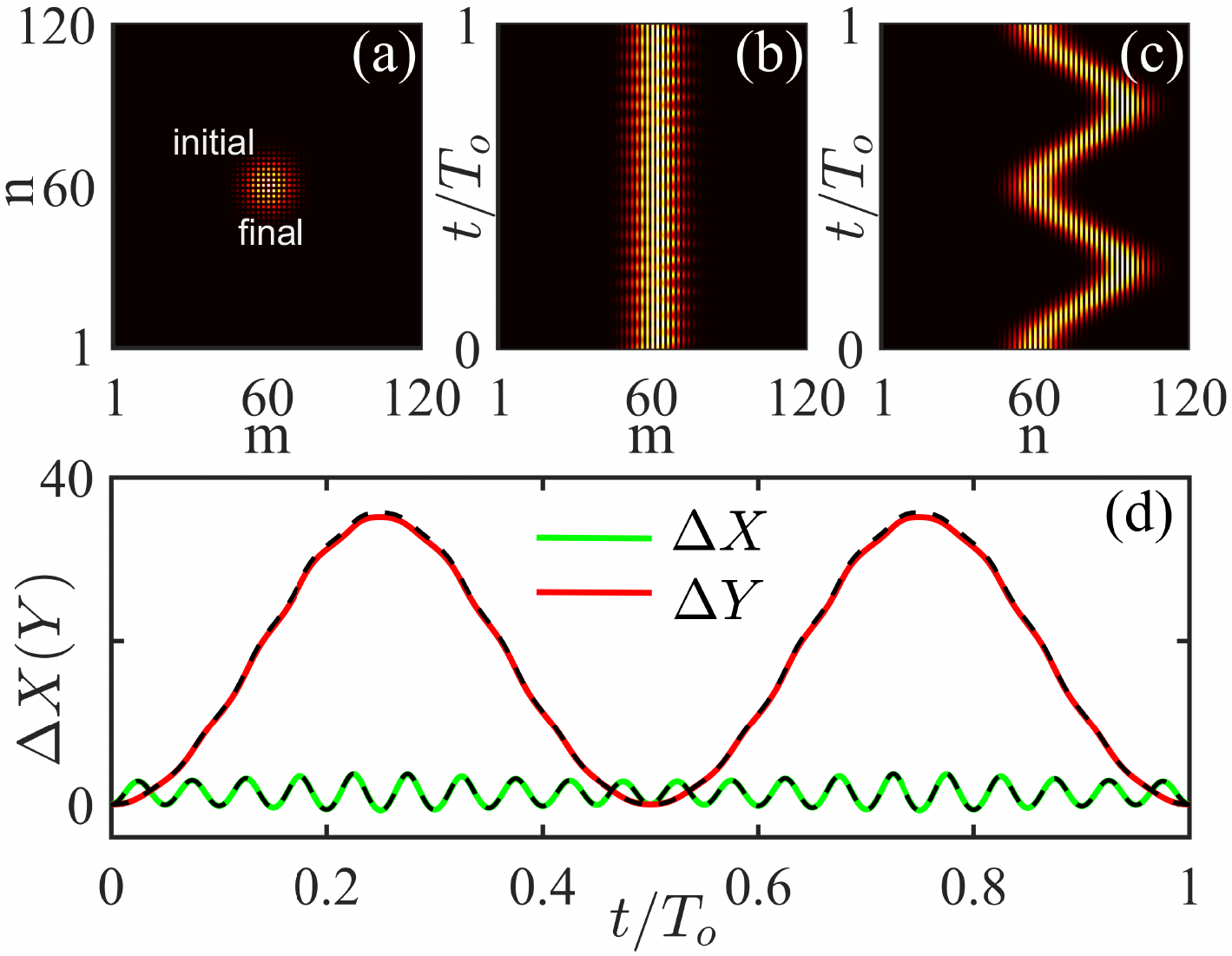}
	\caption{Wavepacket dynamics in topologically trivial phase.
		(a) Density distribution of initial states $(t=0)$ and final states $(t=T_{o})$ in real space.
		(b) and (c) correspond to the density evolution projected to $x$ and $y$ directions, respectively.
		(d) Drift $\Delta X$ and $\Delta Y$ versus time $t$.
		The green and red solid lines are obtained from quantum dynamics, and the black dashed line is obtained via Eq.~\eqref{Ham_6}.
		The other parameters are chosen as $\tau_{x}=\tau_{y}=0.2$, $D=6$, $\eta_x/\eta_y=10$, $F_x=0.04, F_y=0.004$, $\delta=4\tau_{x}$ and $k_{x_0}=k_{y_0}=0$.
	}\label{Fig6}
\end{figure}

When a particle in a seperable band is driven by tilted potentials, the adiabatic condition can be easily satisfied if the tilted potentials are weak enough.
The displacement can be predicted by the semiclassical formula based on the theorem of adiabatic transport.
We first consider the quantum dynamics of a Gaussian wavepacket in the first band which is topologically nontrivial.
The parameters are set as $\tau_{x}=\tau_{y}=0.2$, $D=6$ and $\delta=0$, which correspond to the first band with Chern number $C_1=1$.
$F_x=0.04, F_y=0.004$ are chosen according to the method proposed above, so that the large ratio $\eta_x/\eta_y=10$ supports well quantization of displacement and meanwhile the small $F_x$ and $F_y$ can ensure well adiabatic dynamics.
Fig.~\ref{Fig5} shows the density distributions of initial state $(t=0)$ and final state $(t=T_{o})$ in real space, the time-evolution of wavefunction and the drift of the wavepacket center as a function of time along $x$ and $y$ directions.

First, the wavepacket always maintains well spatial localization in an overall period, see Fig.~\ref{Fig5}(a).
By tracing its dynamical evolution along $x$ and $y$ directions,
the wavepacket is shifted rightward in $x$ direction
and lefttoward in $y$ direction, see Figs.~\ref{Fig5}(b) and \ref{Fig5}(c).
For clarity, the drifts $\Delta X$ and $\Delta Y$ versus time $t$ are given in Fig.~\ref{Fig5}(d).
We can see that
the results obtained by the semi-classical expressions of Eq.~\eqref{Ham_6}
agree well with the ones directly obtained by wavepacket dynamical calculation.
Especially, at  $t=T_{o}$, we can obtain $\Delta X(T_{o})=4$ and $\Delta Y(T_{o})=-40$,
and the RCNs $C_{1,x}^0/\eta_y=-C_{1,y}^0/\eta_x=1$ by combining Eqs.~\eqref{Ham_7} and \eqref{Ham_10},
which are consistent with the first band Chern number $C_1=1$.
By contrast, for the topologically trivial phase, the wavepackets
of the initial state at $(t=0)$
and the final state at $(t=T_{o})$ in real space also maintain spatial localization,
and the density distributions of the initial and final states are overlapped in space. It means that the final state comes back to the position of the initial state, see Fig.~\ref{Fig6}(a).
The Bloch oscillations of the wavepacket in $x$ and $y$ directions have different
periods and amplitudes, as shown in Figs.~\ref{Fig6}(b) and \ref{Fig6}(c).
Indeed, at $t=T_{o}$, one can obtain $\Delta X(T_{o})=0$ and $\Delta Y(T_{o})=0$,
and the RCNs $C_{1,x}^0/\eta_y=-C_{1,y}^0/\eta_x=0$ are consistent with the first band Chern number $C_1=0$, see Fig.~\ref{Fig6}(d).
To verify the mean quasimomentum of an initial Gaussian wavepacket  can be arbitrarily chosen, we also give the drift $\Delta X$ and $\Delta Y$ versus time $t$ for different ($k_{x_0}, k_{y_0}$) in Appendix~\ref{Appendix3}, the results are consistent with the theoretical prediction.
Therefore, we can directly measure band topology via the quantized drift of Bloch oscillations.

\subsection{Direct measurement of super-band topology} \label{super-band}
In this subsection, we will show the possibility of measuring band topology for a composite super-band which consists of two bands touching at several degenerate points.
Since the initial state is mostly localized at momentum $(k_{x_0},k_{y_0})$ of the $l$th band, we can better understand the dynamics in the momentum space.
According to Eq.~\eqref{Ham_3},  in the four-band systems the momentum $(\kappa_x, \kappa_y)=(k_{x_0}-F_xt,k_{y_0}-F_y t)$ is linearly driven by a constant force $(\dot{\kappa}_x, \dot{\kappa}_y)=-(F_x, F_y)$~\cite{2015Geometry}.
If the constant force is large enough, Landau-Zener transitions will always happen especially near anticrossing points of energy bands.
We characterize the Landau-Zener transition by the transition probability from $l$th to $l'$th bands,
\begin{equation}
P_{l,l'}(t)=|\langle \psi(t)| \mu_{l'}(k_x,k_y,t)\rangle|^2,
\end{equation}
where $l,l'=1,2,3,4$ and $|\mu_{l'}(k_x,k_y,t)\rangle$ is the $l'$th eigenstate of $h(k_x,k_y,t)$.
To suppress Landau-Zener transition, the system needs to take a shorter time to travel across the anti-crossing point
than the Zener tunneling time $T_{lz}=\sqrt{\xi} \max(1, \sqrt{\xi})/\Delta$, where $\xi=\Delta^2/(4\sqrt{F_x^2+F_y^2})$ is the adiabaticity
parameter and $\Delta$ is the minimal energy gap~\cite{2010Landau,2020Nonlinear}.
Therefore, for an isolated band, due to a finite energy gap $\Delta\neq 0$,
one can always find proper weak force ($F_x, F_y$) that makes the system evolve adiabatically.
For the composite super-band, it seems impossible to satisfy the adiabatic condition for individual bands because there exist gap closing points.
Indeed, it is true for the conventional scheme with a tilted field in only one direction such as $x$-direction, in which the initial state needs to uniformly occupy the band involving all $k_y$.
Consequently, Landau-Zener transition is unavoidable when sweeping the gap closing points under the tilted force in $x$-direction.
However, based on our scheme with tilts in both $x$ and $y$ directions, there is no requirement of uniform band occupation.
The trajectory of $(\kappa_x, \kappa_y)$ may avoid the degenerated points in two-dimensional Brillouin zone by selecting the appropriate initial momentum $(k_{x_0}, k_{y_0})$, and hence it is possible to measure the Chern number for an energy-inseparable super-band via the RCN.

\begin{figure}[htp]
\center
\includegraphics[width=\columnwidth]{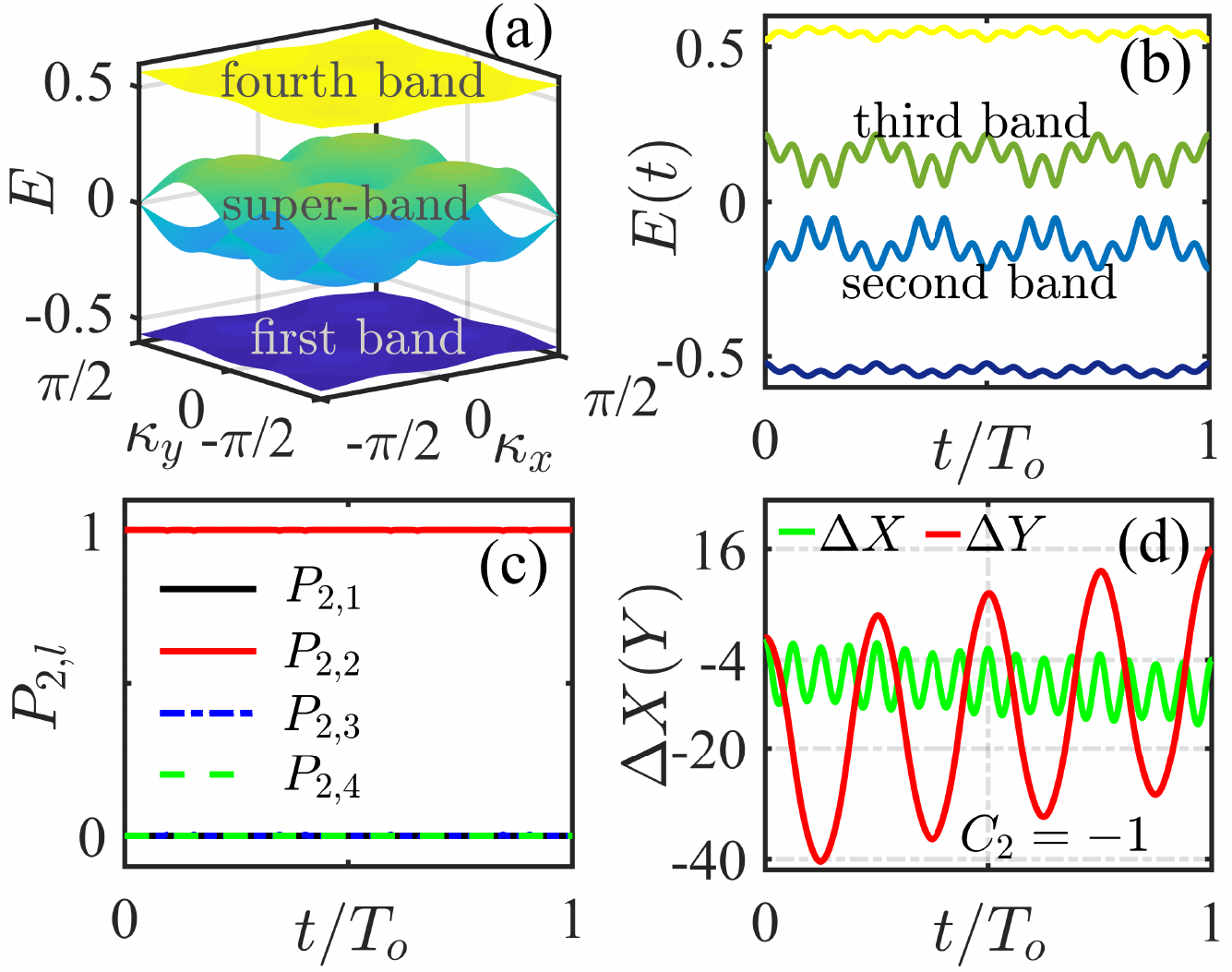}
\caption{(a) Energy spectrum for the Harper-Hofstadter Hamiltonian.
(b) Eigenvalues change with the trajectory $(\kappa_x(t), \kappa_y(t))$.
(c) Transition probability for Landau-Zener tunneling as a function of evolution time $t$.
(d) Drift $\Delta X$ and $\Delta Y$ versus time $t$.
The other parameters are chosen as $\tau_{x}=\tau_{y}=0.2$, $\eta_x/\eta_y=4$, $F_x=0.002, F_y=0.0005$, $\delta=0$ and $k_{x_0}=k_{y_0}=\pi/4$.
}\label{Fig7}
\end{figure}

To be clear, we consider such a system with super-band by choosing the parameters as $\tau_{x}=\tau_{y}=0.2$ and $\delta=0$,
whose energy bands are shown in Fig.~\ref{Fig7}(a).
We find that the second and third bands form a composite super-band with energy degeneracies at some momenta.
Such super-band is well separated from the first and fourth energy bands.
The Chern numbers of individual bands in the super-band are not well-defined due to the difficulty in separating the two bands at the degenerated points.
A common Chern number has been used to describe the overall topology
of the composite super-band~\cite{Nakahara2018},
\begin{equation}
C_{s}= \frac{1}{{2\pi i }}\int_{BZ} d k_y{dk_x \textrm{Tr} [\mathcal{F}({k_x,k_y})]}.
\label{Chern}
\end{equation}
Here, $[\mathcal{F}({k_x,k_y})]=\partial_{k_x} A_{k_y}-\partial_{k_y} A_{k_x} + i[A_{k_x},A_{k_y}]$ is the non-Abelian Berry curvature with Berry connection elements  $[A_\mu]_{m,n}=\langle \psi_{m}|\nabla_\mu|\psi_{n}\rangle$, where $m,n$ are the band indices in the super-band.
When the super-band only includes a single band, because the commutator vanishes, the non-Abelian generalization of Chern number is reduced to the conventional one [Eq.~\eqref{Chern}].
Even when the super-band contains more gapless bands with finite degenerated points, if we carefully exclude the Berry curvature of the degenerated points, the Chern number receives contribution only from the Abelian part of Berry curvatures~\cite{Mochol_Grzelak_2018}.
Alternatively, the Chern number of the super-band is taken as if the sum of effective ``Chern numbers" (excluding the Berry curvature of degenerated points) of the second and third bands.
Eq.~\eqref{Chern} gives the same value of $C=-1$ for the second and third bands excluding the degenerated points.
The Chern numbers $C_1, C_s, C_4$ corresponding to the first band, super-band and the fourth band take the values $\{1,-2,1\}$,
respectively.
We should further emphasize that the Chern number of the super-band $C_s$ can not be directly measured by conventional Thouless pumping~\cite{ke2016topological}
or using the tilt in only one direction~\cite{aidelsburger2015measuring}, where all momenta are equally involved. This is because these methods will lead to non-adiabatic transitions near
the degenerated points, and then hinder the measurement of topological invariants.

However, the RCN only depends on the trajectory of $(\kappa_x(t), \kappa_y(t))$ in two-dimensional Brillouin zone.
By choosing appropriate initial momentum $(k_{x_0}, k_{y_0})$ and tilted fields,
one can ensure that energy gaps of the four energy bands remain open along the trajectory $(\kappa_x(t), \kappa_y(t))$,
where the super-band can be decomposed to two separable energy bands (i.e. the second and third bands), as shown in Fig.~\ref{Fig7}(b).
%
Without lossing of generality, we show how to measure RCN of the second band by preparing initial state in the second band,
$|\psi(t_0)\rangle=|\mu_2(k_{x_0}, k_{y_0}, 0)\rangle$, with initial momentum $(k_{x_0}, k_{y_0})=(\pi/4, \pi/4)$.
To ensure that the system is adiabatically evolving, we choose weak tilted fields as $F_x=0.002$ and  $F_y=0.0005$.
Fig.~\ref{Fig7}(c) shows the transition probability from the second band to the $l$th band, $P_{2,l}$, along the trajectory $(\kappa_x(t), \kappa_y(t))$.
It is clear that the transition probability $P_{2,2}$  is always equal to one,
which means that the system is adiabatically following the eigenstates in the second band.
Then, the drifts $\Delta X$ and $\Delta Y$ versus time $t$ are given in Fig.~\ref{Fig7}(d).
At $t=T_{o}$, we can obtain $\Delta X(T_{o})=-4, \Delta Y(T_{o})=16$,
and the RCN $C_{2,x}^0/\eta_y=-C_{2,y}^0/\eta_x=-1$ by using Eq.~\eqref{Ham_10}.
%
Similarly, by choosing the initial state $|\psi(t_0)\rangle=|\mu_3(k_{x_0}, k_{y_0}, 0)\rangle$,
we also can obtain the RCN $C_{3,x}^0/\eta_y=-C_{3,y}^0/\eta_x=-1$, as a direct measurement of the effect ``Chern number" excluding the degenerated points.
%
Then, the Chern number of the super-band ($C_s=-2$) is consistent with the summation of reduced Chern numbers of the second and third bands, $(C_{2,x}^0+C_{3,x}^0)/\eta_y=-2$.
Therefore, we provide a direct method to measure the Chern number of the super-band via the quantized drift in two-dimensional Bloch oscillations.

\section{Conclusion and discussion \label{SSec5}}
In summary, we reveal an almost perfectly quantized drift of two-dimensional Bloch oscillations
in a topological Chern insulator. The quantized drift is related to a reduced Chern number defined by line integral of Berry curvature in each direction, as an effective measurement of the conventional Chern number.
Our scheme can apply to extracting Chern number of both energy-separable and energy-inseparable bands.
Compared with schemes of Thouless pumping or integer quantum Hall effect,
our scheme does not require equal superposition of states involving all momenta,
but any particular momentum state in a band can be chosen as the initial state.
Meanwhile, the introduction of tilts also well suppresses the diffusion of Gaussian wavepackets
and improves the measurement accuracy of the Chern number near the phase transition point.

In this work, we show how to use Bloch oscillations to extract Chern number of a super-band even when there exist several degenerate points.
This is because the Bloch oscillations in the momentum space avoid the degenerate points and the non-Abelian part of the Berry curvature takes no effect.
If the super-band contains completely degenerated bands, there is no way to avoid the degeneracy and non-Abelian Berry curvature must play an important role in the wavepacket dynamics.
It is unclear how to apply our method to measure topological invariants such as second Chern number of a completely degenerated super-band~\cite{Sugawa1429,Mochol_Grzelak_2018}.
Inversely, it is also very interesting to explore how to tail the Bloch oscillations with such nontrivial topological super-band.

%
Recently, we became aware of two new works studying the center-of-mass drift in a fractional Chern insulator subjected to a tilted field in only one direction~\cite{2020Detecting,2020Fractional}.
However, the drifts are not quite well quantized, which may come from the fluctuation of Berry curvature in different center-of-mass momenta.
If the tilted fields are introduced along both directions, the center-of-mass momentum may be sampled in a more uniform way and the fluctuation may be wiped away.
We believe our method can be extended to interacting topological systems~\cite{Rachel2018} and could potentially provide more accuracy measurement of fractional topological states.

\begin{acknowledgments}
The authors are thankful to Chaohong Lee for enlightening suggestions and helpful discussions.
This work is supported by the National Natural Science Foundation of China under Grant No.11805283, the Hunan Provincial Natural Science Foundation under Grant No.2019JJ30044, the Scientific Research Fund of Hunan Provincial Education Department under Grant No.19A510 and the Talent project of Central South University of Forestry and Technology under Grant No.2017YJ035.
Y.K. is supported by the National Natural Science Foundation of China (Grant No.11904419).
\end{acknowledgments}

\appendix

\section{The Hamiltonian in momentum space } \label{Appendix1}
According to the Hamiltonian~\eqref{Ham_2}, the eigenvalue equation can be written as,
\begin{eqnarray} \label{Ham_13}
-E|m,n\rangle&=&\tau_{x} e^{i F_{x} t}|m+1,n\rangle+\tau_{y} e^{i F_{y} t} e^{i 2 \pi \beta m}|m,n+1\rangle \nonumber \\
&+&\tau_{x} e^{-i F_{x} t}|m-1,n\rangle+\tau_{y} e^{-i F_{y} t} e^{-i 2 \pi \beta m}|m,n-1\rangle \nonumber \\
&+&\frac{\delta}{2} \left[(-1)^{m}+(-1)^{n}\right]|m,n\rangle.
\end{eqnarray}
Here we consider $\beta=1/4$ which describes hopping on the square lattice in the presence
of a magnetic flux $2 \pi \beta= \pi/2$ per plaquette.
To solve this equation, we make the following ansatz for the wave function:
\begin{eqnarray} \label{Ham_14}
|m,n\rangle=e^{i k_{x} m} e^{i k_{y} n}\left\{\begin{array}{ll}
\psi_{A}, & \text { for } m, n \text { odd }; \nonumber \\
\psi_{B} e^{i m \pi / 2} & \text { for } m \text { even }, n \text { odd }; \nonumber \\
\psi_{C} & \text { for } m \text { odd, } n \text { even }; \nonumber \\
\psi_{D} e^{i m \pi / 2} & \text { for } m, n \text { even }.
\end{array}\right.
\end{eqnarray}
Here, $k_x, k_y$ are defined within the first magnetic Brillouin zone.
Inserting this ansatz into the Schr\"{o}dinger equation we obtain the following $4\times4$ eigenvalue equation
\begin{eqnarray} \label{Ham_15}
h(k_x,k_y,t)\left(\begin{array}{l}
\psi_{A} \\
\psi_{B} \\
\psi_{C} \\
\psi_{D}
\end{array}\right)=E(k_x,k_y,t)\left(\begin{array}{c}
\psi_{A} \\
\psi_{B} \\
\psi_{C} \\
\psi_{D}
\end{array}\right),
\end{eqnarray}
with
\begin{eqnarray}
h=2\left(\begin{array}{cccc}
-\frac{\delta}{2} & -\tau_{y}\cos \kappa_{y} & -\tau_{x}\cos \kappa_{x} & 0 \\
-\tau_{y}\cos \kappa_{y} & 0 & 0 & \tau_{x}\sin \kappa_{x} \\
-\tau_{x}\cos \kappa_{x} & 0 & 0 & i \tau_{y}\sin \kappa_{y} \\
0 & \tau_{x}\sin \kappa_{x} & -i \tau_{y}\sin \kappa_{y} & \frac{\delta}{2}
\end{array}\right), \nonumber \\
\end{eqnarray}
where $\kappa_x=k_x-F_x t$, $\kappa_y=k_y-F_y t$, $k_x\in[-\pi/2,\pi/2]$ and $k_y\in[-\pi/2,\pi/2]$.

\section{Relation between reduced Chern number and Chern number } \label{Appendix2}

When the tilt is absent, the model~\eqref{Ham_1} is reduced to the Harper-Hofstadter-like topological model,
$\widetilde{H}=H_1+H_2$,
whose Hamiltonian in momentum space is given by $\widetilde{h}(k_x,k_y)=h(k_x, k_y, 0)$.
Compared to the case with tilts,
the corresponding bands and eigenstates can be obtained by replacing $\kappa_x$ by $k_x$ and $\kappa_y$ by $k_y$.
The conventional Chern number can be defined by the integral of Berry curvature in the two-dimensional Brillouin zone as
\begin{eqnarray} \label{Ham_9s1}
C_{l}=\frac{1}{2 \pi} \int_{-\pi / 2}^{\pi / 2} d k_{x} \int_{-\pi/2}^{\pi/2} d k_{y} \mathcal{F}_{l} (k_{x}, k_{y}),
\end{eqnarray}
where
\begin{equation}
\mathcal{F}_{l} (k_{x}, k_{y})=-2 \operatorname{Im}\sum_{l^{\prime} \neq l} \frac{\left\langle u_{l}\left|\partial_{k_x} \widetilde{h} \right| u_{l^{\prime}}\right\rangle\left\langle u_{l^{\prime}}\left|\partial_{k_y} \widetilde{h}\right| u_{l}\right\rangle}{\left(E_{l}-E_{l^{\prime}}\right)^{2}}.
\end{equation}
The Chern number can be further written as
\begin{eqnarray}
C_l&&=\frac{1}{2 \pi} \int_{-\pi / 2}^{\pi / 2} d k_{x} \int_{-\pi/2}^{\pi/2} d \kappa_{y} \mathcal{F}_{l} (k_{x}, \kappa_{y}) \nonumber \\
&&=\frac{1}{4 \pi} \int_{-\pi / 2}^{\pi / 2} d k_x \int_{0}^{T_{y}} \mathcal{F}_{l,x}(k_x,k_y,t) d t \nonumber \\
&&=\frac{1}{\eta_y \pi} \int_{-\pi / 2}^{\pi / 2} d k_x C_{l,x}(k_{x}, k_{y}).
\label{Chern}
\end{eqnarray}
When $\eta_x/\eta_y$ approaches $0$ or $\infty$, $C_{l,x}(k_{x}, k_{y})$ tends to be an integer $C_{l,x}^0$, almost independent of both $k_x$ and $k_y$ (see Appendix~\ref{Appendix4} for details). Hence we can get rid of the average over $k_x$ in the above equation and yield $C_l=C_{l,x}^0/\eta_y$.
In the similar way, the Chern number can be also written as
\begin{eqnarray} \label{Ham_9s2}
C_{l}=-\frac{1}{\eta_x \pi} \int_{-\pi / 2}^{\pi / 2} d k_y C_{l,y}(k_{x}, k_{y})=-C_{l,y}^0/\eta_x.
\end{eqnarray}
As a consequence, the Chern number is related to  $C_{l,x}^0$ and $C_{l,y}^0$ via
\begin{eqnarray}
C_{l} = C_{l,x}^0/\eta_y=-C_{l,y}^0/\eta_x. \label{ClCxCy}
\end{eqnarray}
Indeed, $C_{l,x}$ and $C_{l,y}$ defined by a one dimensional integral can be
effectively regarded as a reduced expression for the  Chern number $C_{l}$,
which we call \textit{Reduced  Chern  number}(RCN)~\cite{Kepumping2020}.

\section{Convergence of reduced Chern number when $F_x/F_y=\eta_x/\eta_y$ approaches $0$ or $\infty$} \label{Appendix4}

When $F_x/F_y=\eta_x/\eta_y$ approaches $0$ or $\infty$, $C_{l,x}(k_x,k_y)$ and $C_{l,y}(k_x,k_y)$ respectively tend to be integer numbers $C_{l,x}^0$ and $C_{l,y}^0$, independent of the initial momentum value of a Bloch state.
It is mainly because the Bloch oscillations sample the Berry curvature in an uniform way, and the sampling becomes more uniform when $\eta_x/\eta_y$ tends to be $0$ or $\infty$.
%
%
Without loss of generality, we will explain why $C_{l,x}(k_x,k_y)$  are convergent to a constant RCN in the limit of $\eta_x/\eta_y\rightarrow \infty$.

We start from the property of Berry curvatures, that is, $\mathcal F_l (k_x,k_y,t)$  is a periodic function of quasi-momentum and time. Thus, the integral of Berry curvature over an overall period of time is independent of the initial time,
\begin{eqnarray}
{C_{l,x}}({k_x},{k_y}) & =& \frac{1}{q}\int_0^{{T_o}} {{{\cal F}_{l,x}}({k_x},{k_y},t)dt}
\label{ChernRxy} \\
& =& \frac{1}{q}\int_0^{{T_o}} {{{\cal F}_{l,x(y)}}({k_x},{k_y},t - \Delta {k_x}/{F_x})dt}.  \nonumber
\end{eqnarray}
We know that the quasi-momentum  $({\kappa _x},{\kappa _y})$ change with time as ${\kappa _{x(y)}} = {k_{x(y)}} - {F_{x(y)}}t$.
When we shift the time to $t - \Delta {k_x}/{F_x}$, it is equivalent to shift $k_x$  to ${k_x} + \Delta {k_x}$   and $k_y$  to ${k_y} + \Delta {k_y}$  while maintain the time as $t$.
Eq.~\eqref{ChernRxy} can be further written as
\begin{eqnarray}
&&\frac{1}{q}\int_0^{{T_o}} {{{\cal F}_{l,x(y)}}({k_x},{k_y},t - \Delta {k_x}/{F_x})dt}   \\
&=& \frac{1}{q}\int_0^{{T_o}} {{{\cal F}_{l,x(y)}}({k_x} + \Delta {k_x},{k_y} + {F_y}/{F_x}\Delta {k_x},t)dt}.  \nonumber
\end{eqnarray}
It means that
\begin{eqnarray}
{C_{l,x}}({k_x},{k_y}) = {C_{l,x}}({k_x} + \Delta {k_x},{k_y} + {F_y}/{F_x}\Delta {k_x}).
\end{eqnarray}
When $F_x/F_y=\eta_x/\eta_y\rightarrow \infty$, for arbitrary $\Delta k_x$, we can make an approximation,
\begin{eqnarray}
{C_{l,x}}({k_x},{k_y}) &=& {C_{l,x}}({k_x} + \Delta {k_x},{k_y} + {F_y}/{F_x}\Delta {k_x})  \nonumber\\
&\approx& {C_{l,x}}({k_x} + \Delta {k_x},{k_y}). \label{Clxa}
\end{eqnarray}
Making $\Delta k_x={F_x}/{F_y}\Delta {k_y}$, one can have
\begin{eqnarray}
{C_{l,x}}({k_x},{k_y}) \approx {C_{l,x}}({k_x} + {F_x}/{F_y}\Delta {k_y},{k_y}). \label{Clxa2}
\end{eqnarray}

Similarly, Eq.~\eqref{ChernRxy} also can be written as
\begin{eqnarray}
&&\frac{1}{q}\int_0^{{T_o}} {{{\cal F}_{l,x(y)}}({k_x},{k_y},t - \Delta {k_y}/{F_y})dt}   \\
&=& \frac{1}{q}\int_0^{{T_o}} {{{\cal F}_{l,x(y)}}({k_x} + {F_x}/{F_y}\Delta {k_y},{k_y} + \Delta {k_y},t)dt}.  \nonumber
\end{eqnarray}
It means that
\begin{eqnarray}
{C_{l,x}}({k_x},{k_y}) = C({k_x} + {F_x}/{F_y}\Delta {k_y},{k_y} + \Delta {k_y}). \label{Clxb2}
\end{eqnarray}
Combining Eqs.~\eqref{Clxa2} and \eqref{Clxb2}, One can immediately obtain
\begin{equation}
{C_{l,x}}({k_x'},{k_y}) \approx {C_{l,x}}({k_x'},{k_y}+\Delta k_y).
\end{equation}
for any arbitrary $\Delta k_y$.
From Eqs.~\eqref{Clxa} and \eqref{Clxb2}, it is found that $C_{l,x}(k_x,k_y)$ are almost the same and denoted as $C_{l,x}^0$, independent of both $k_x$  and $k_y$  in the limit of $\eta_x/\eta_y\rightarrow \infty$. The Chern number is related to the RCN as
\begin{equation}
{C_l} = \frac{1}{{{\eta _y}\pi }}\int_{ - \pi /2}^{\pi /2} d {k_x}{C_{l,x}}({k_x},{k_y}) = \frac{{C_{l,x}^0}}{{{\eta _y}}}.
\end{equation}
By swapping $k_x$  and $k_y$, we can also prove the relation Eq.~\eqref{ClCxCy}   in the limit of $\eta_x/\eta_y$ approaches $0$ or $\infty$.

\begin{figure}[!htp]
	\center
	\includegraphics[width=\columnwidth]{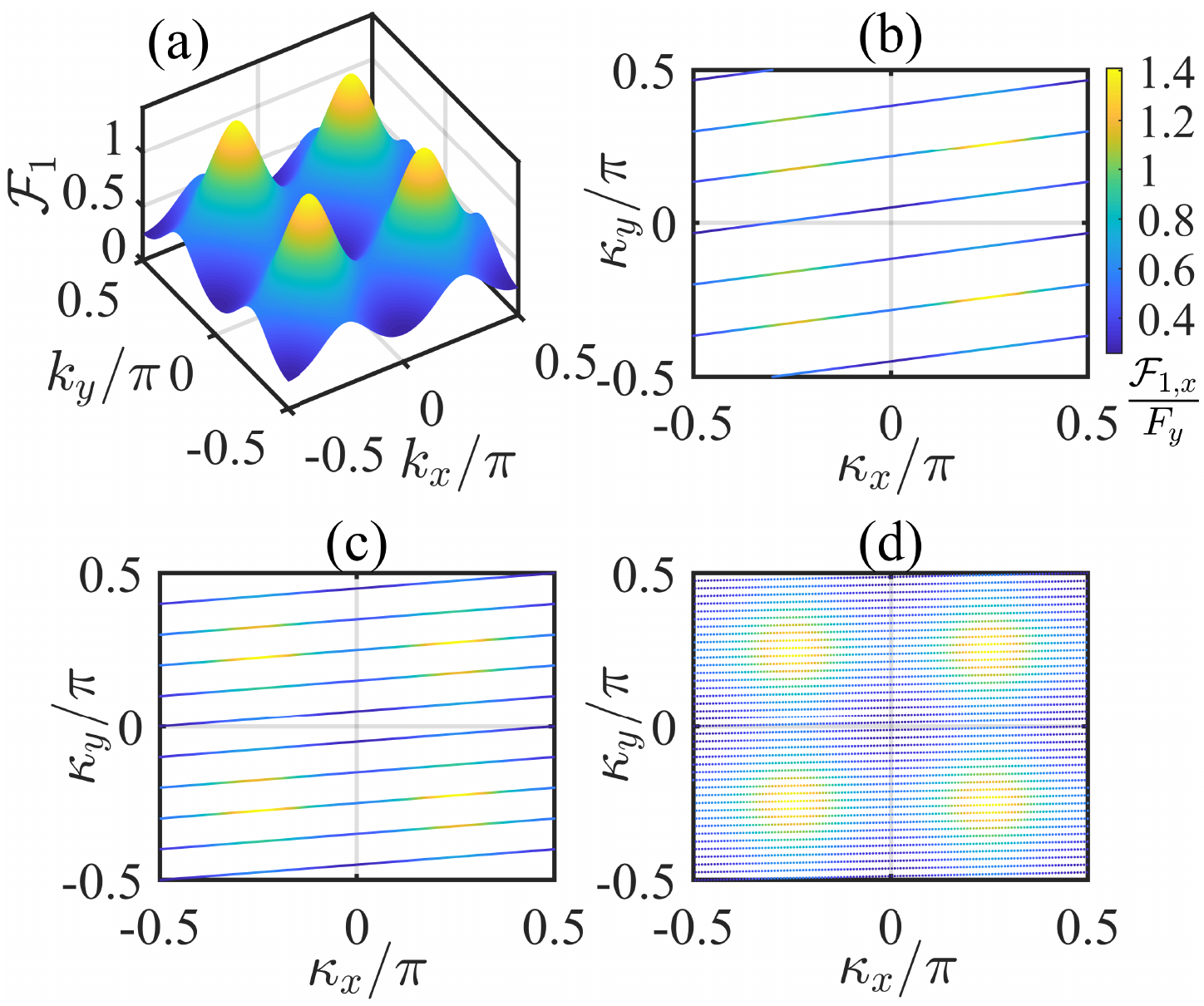}
	\caption{ (a) The distribution of Berry curvature $\mathcal{F}_{1}\left(k_{x}, k_{y}\right)$ in two-dimensional Brillouin zone. (b)-(d) Sampling trajectories of ${\mathcal{F}_{1, x}}/{F_{y}}$ respectively correspond to $F_x/F_y=\eta_x/\eta_y=6$, $10$ and $40$ with fixed norm $\sqrt{F_x^2+F_y^2}=0.0402$. The other parameters are chosen as $\tau_{x}=\tau_{y}=0.2, \delta=0$.
	}\label{Fig8}
\end{figure}

As an example, we give the distribution of Berry curvature $\mathcal{F}_{1}\left(k_{x}, k_{y}\right)$ in the  two-dimensional Brillouin zone; see Fig.~\ref{Fig8}(a).
Fig.~\ref{Fig8}(b)-(d) correspond to sampling trajectories of ${\mathcal{F}_{1, x}}/{F_{y}}$  in the momentum space for $\eta_x/\eta_y=6$, $10$ and $40$, respectively.
We clearly see that the sampling density increases with the increase of $\eta_x/\eta_y$.
When the sampling density is large enough, the two-dimensional integral about the conventional Chern number can be replaced by the linear integral of sampling trajectory.
Because the initial momentum value only determines the initial position of the sampling trajectory, without affecting sampling density,
we have $C_{1, x}\left(k_{x}, k_{y}\right) / \eta_{y}=C_{l, x}^{0} / \eta_{y}$.
We have to emphasize that the almost quantized RCN is quite general and it does not require the uniform distribution of Berry curvature.

\begin{figure}[htp]
	\center
	\includegraphics[width=\columnwidth]{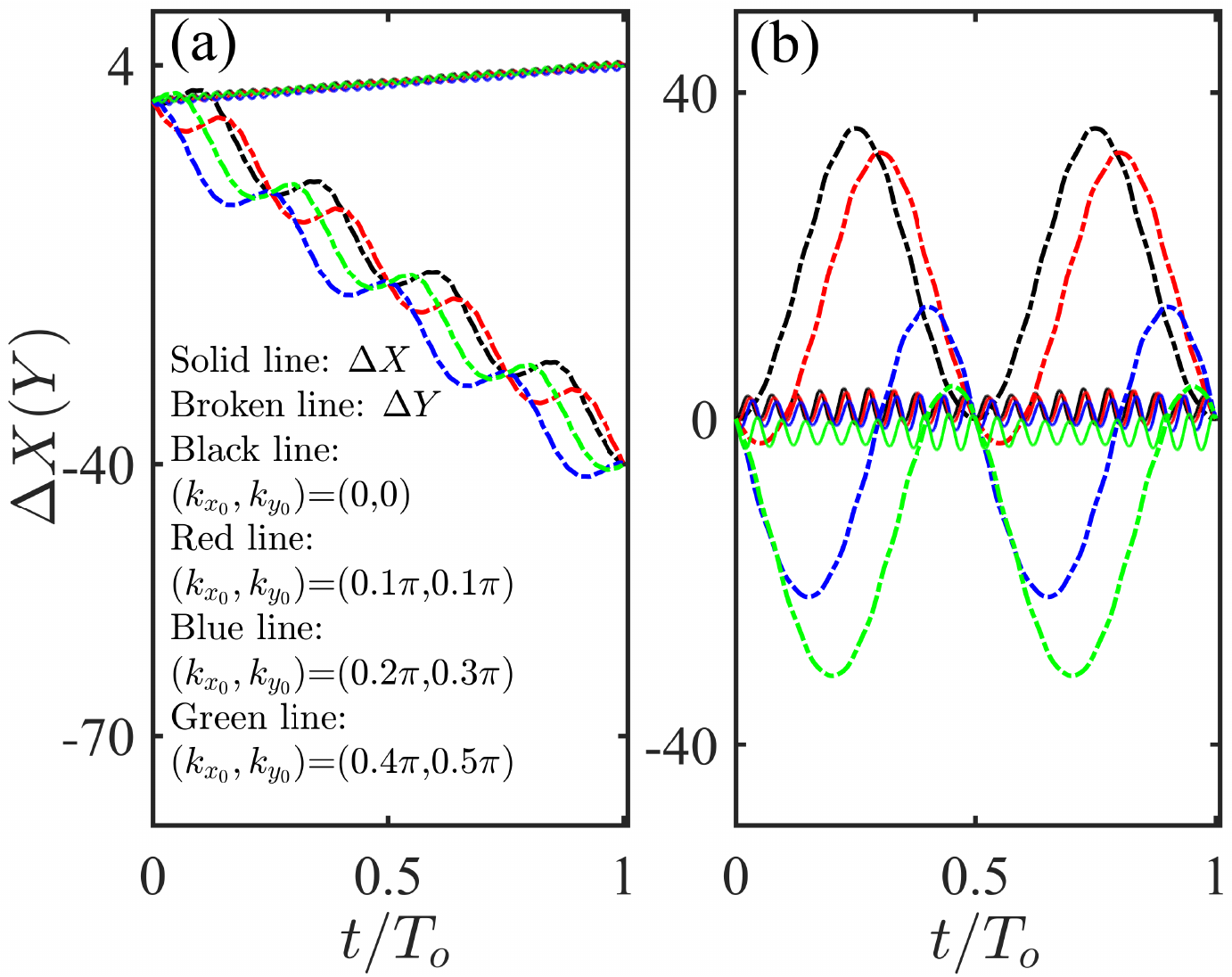}
	\caption{Drift $\Delta X$ and $\Delta Y$ versus time $t$ for different ($k_{x_0}, k_{y_0}$).
		(a) Topologically nontrivial phase ($C_1=1$, $\delta=0$). (b)Topologically trivial phase ($C_1=0$, $\delta=4\tau_{x}$). The other parameters are chosen as $\tau_{x}=\tau_{y}=0.2$, $\eta_x/\eta_y=10$, $F_x=0.04$ and $F_y=0.004$.}\label{Fig9}
\end{figure}

\section{The Influence of initial quasimomentum on quantized drift} \label{Appendix3}

To verify that the mean quasimomentum of an initial Gaussian wavepacket  can be arbitrarily chosen, we give the drift $\Delta X$ and $\Delta Y$ versus time $t$ for different initial quasimomentum ($k_{x_0}, k_{y_0}$) in Fig.~\ref{Fig9}.
Despite the drift trajectories of $\Delta X$ and $\Delta Y$ have a significant difference for different mean quasimomentum, but these trace lines will converge at one point when $t=T_{o}$.
Fig.~\ref{Fig9}(a) corresponds to the topologically nontrivial phase, no matter how we choose the initial mean quasimomentum, we always obtain $\Delta X(T_{o})=4, \Delta Y(T_{o})=-40$,
and the RCNs $C_{1,x}^0/\eta_y=-C_{1,y}^0/\eta_x=1$ by combining the Eqs.~\eqref{Ham_7} and \eqref{Ham_10},
which are consistent with the first band Chern number $C_1=1$.
Similarly, one can also obtain $\Delta X(T_{o})=0, \Delta Y(T_{o})=0$ in Fig.~\ref{Fig9}(b),
and the RCNs $C_{1,x}^0/\eta_y=-C_{1,y}^0/\eta_x=0$ are consistent with the first band Chern number $C_1=0$.
These results prove again that our method is rather insensitive and robust to the choice of initial quasimomentum.

\bibliography{quantum_quench}

\end{document}